\documentclass[notitlepage, amssymb,amsmath,aps,floatfix,prd,nofootinbib,superscriptaddress,showpacs,preprintnumbers]{revtex4-1}

\usepackage[utf8]{inputenc}

\usepackage{natbib}
\usepackage{graphicx}
\usepackage{epsfig}
\usepackage{amsmath}
\input{epsf}
\usepackage{psfrag}

\usepackage[usenames,dvipsnames]{color}

\newcommand{\beq}{\begin{eqnarray}}
\newcommand{\eeq}{\end{eqnarray}}

\newcommand{\nn}{\nonumber \\}

\usepackage{amsmath,color}
\usepackage{amssymb}
\usepackage{bm}
\usepackage{graphicx}
\usepackage{multirow}

\begin{document}
\preprint{LA-UR-24-22492}

\title{Spin-orbit entanglement in the Color Glass Condensate }

\author{Shohini Bhattacharya}
\email{shohinib@lanl.gov}
\affiliation{Theoretical Division, Los Alamos National Laboratory, Los Alamos, New Mexico 87545, USA}

\author{Renaud Boussarie}
\email{renaud.boussarie@polytechnique.edu}
\affiliation{CPHT, CNRS, Ecole Polytechnique, Institut Polytechnique de Paris, 91128 Palaiseau, France}

\author{Yoshitaka Hatta}
\email{yhatta@bnl.gov}
\affiliation{Physics Department, Brookhaven National Laboratory, Upton, NY 11973, USA}
\affiliation{RIKEN BNL Research Center, Brookhaven National Laboratory, Upton, NY 11973, USA}

\begin{abstract}
 We compute the spin-orbit correlations of quarks and gluons at small-$x$ and show that the helicity and the orbital angular momentum  of individual partons are strongly  anti-aligned even in unpolarized or spinless hadrons and nuclei. Combined with the fact that gluons in the Color Glass Condensate are linearly polarized, our finding indicates that the helicity and the orbital angular momentum of single gluons are maximally entangled in a quantum mechanical sense.

\end{abstract}

\maketitle

\section{Introduction}

One of the core missions of the future Electron-Ion Collider (EIC) experiment  \cite{AbdulKhalek:2021gbh} at Brookhaven National Laboratory is to unambiguously discover and then explore the phenomenon of the gluon saturation \cite{Gribov:1983ivg,Gelis:2010nm,Morreale:2021pnn}.
that universally occurs  in all hadrons and nuclei alike when they are boosted to sufficiently high energies. 
This was clearly expressed in the influential National  Academy of Science report \cite{nas} which concluded  that the EIC can uniquely answer the profound question {\it ``What are the emergent properties of dense systems of gluons?"} where `dense systems' refer to gluon saturated matter commonly dubbed the Color Glass Condensate (CGC).  Already a number of remarkable properties of the CGC have been exposed through decades of extensive theoretical  studies \cite{Gribov:1983ivg,Gelis:2010nm,Morreale:2021pnn}.  In this paper we uncover a novel emergent property of the CGC that has been unnoticed so  far: Perfect spin-orbit {\it anti}-correlation for quarks and gluons at small-$x$.

Spin-orbit correlation in the present context concerns the relative orientation of the helicity and orbital angular momentum (OAM) of individual quarks and gluons inside a hadron or a nucleus, similar to the spin-orbit coupling of an electron  in a hydrogen atom. Such a notion was introduced in the nucleon structure physics in   \cite{Lorce:2011kd} and can be quantified by  certain Generalized Transverse Momentum Dependent (GTMD) distributions.  However, despite some progress in the literature   \cite{Lorce:2014mxa,Rajan:2017cpx,Tan:2021osk,Engelhardt:2021kdo,Kim:2024cbq}, at the moment very little  is known about the properties of the correlation from first principles, especially for gluons, let alone their experimental verification \cite{Bhattacharya:2017bvs,Bhattacharya:2018lgm,Boussarie:2018zwg,Bhattacharya:2023hbq}. In this paper, we demonstrate that spin-orbit correlations for both quarks and gluons are  explicitly calculable at small-$x$ in the CGC framework. (See a related work \cite{Boer:2018vdi}.) We find that the correlation not only survives in the eikonal approximation, but also grows strongly with decreasing $x$ in the same manner as the BFKL pomeron   
\cite{Kuraev:1977fs,Balitsky:1978ic} and eventually saturates.    This may  sound surprising at first given that distributions associated with polarization are often labeled as `sub-eikonal' and suppressed by a factor of $x\ll 1$ compared to unpolarized ones (e.g.  \cite{Cougoulic:2022gbk} and references therein). Moreover, the sign of the spin-orbit correlation is consistently negative,  meaning that the helicity and OAM of individual small-$x$ partons  point to opposite directions. This is similar to the known sign difference between the helicity and OAM parton distributions at small-$x$ inside a polarized hadron \cite{Hatta:2016aoc,More:2017zqp,Hatta:2018itc,Boussarie:2019icw,Kovchegov:2023yzd,Manley:2024pcl}. However, the anti-correlation exists even in unpolarized or spinless hadrons. 

Furthermore, we shall argue that `correlation' actually implies `entanglement' between the spin and the OAM of individual gluons in a quantum mechanical sense. 
This provides a novel perspective on the CGC shedding light on the angular momentum characteristics of quarks and gluons at small-$x$. 
 We contrast our finding with the analogous entangled photon states that have been recently observed in the field of quantum optics \cite{stav}.

\section{Gluon spin-orbit correlation} 

For a high energy hadron or nucleus single particle state $|p\rangle$, we define the quark and gluon spin-orbit correlation $C_{q,g}$ as 
\beq
\int \frac{d^3z}{2(2\pi)^3} e^{ixP^+z^--ik_\perp\cdot z_\perp}\langle p'|\bar{\psi}(-z/2)\gamma^+\gamma_5W_{\pm}\psi(z/2)|p\rangle 
&=&  -i\frac{\epsilon_{ij}k_\perp^i \Delta_\perp^j}{M^2}C_q(x,\xi,k_\perp,\Delta_\perp),\label{a3} 
\eeq
\beq
 \frac{i}{x}\int \frac{d^3z}{(2\pi)^3P^+} e^{ixP^+z^--ik_\perp\cdot z_\perp}\langle p'|2{\rm Tr}[W_+\tilde{F}^{+\mu}(-z/2)W_\pm F^{+}_{\ \mu}(z/2)]|p\rangle 
&=& -i\frac{\epsilon_{ij}k_\perp^i \Delta_\perp^j}{M^2}C^{[+\pm]}_g(x,\xi,k_\perp,\Delta_\perp), \label{a4}
\eeq  
where $p^\mu=P^\mu-\Delta^\mu/2$, $p'^\mu = P^\mu+\Delta^\mu/2$ with $P^\mu \approx \delta^\mu_+ P^+$. $\xi=\frac{-\Delta^+}{2P^+}$ is the skewness variable which will be set to zero in the following. 
For definiteness, we have assumed a spinless hadron or nucleus in the above parametrizations. In the case of nonzero spin, additional terms and appropriate spinor/tensor structures should be supplemented. However, such a generalization is not essential for the present purpose. 
$W_{\pm}$ is a staple-shaped Wilson line connecting the two points $\pm z/2$ via the light cone infinity $z^-=\pm \infty$. In the quark case,  the  difference $W_+$ or $W_-$ does not matter. In the gluon case, the two different configurations of $W_{\pm}$ lead to physically different distributions \cite{Bomhof:2006dp} as is well known in the analysis of the unpolarized gluon distributions  \cite{Dominguez:2011wm}. Following  the usual nomenclature, we shall refer to  $C_g^{[++]}$ and $C_g^{[+-]}$  as the Weisz\"acker-Williams (WW) distribution and the dipole distribution, respectively, and use superscripts  `WW' and `dip'  instead of $[+\pm]$.

The existence of the term proportional to $\epsilon_{ij}k_\perp^i\Delta_\perp^j$  was pointed out in  \cite{Meissner:2008ay,Meissner:2009ww}   which was later interpreted as the spin-orbit correlation \cite{Lorce:2011kd}. A quick way to see this is to notice that  $\bar{\psi}\gamma^+\gamma_5\psi$ and $i\tilde{F}^{+\mu}F^+_{\ \mu}$ are the quark and gluon helicity operators,  and the product $-i\epsilon_{ij}k^i_\perp \Delta_\perp^j \sim \vec{b}_\perp\times \vec{k}_\perp$ with $b_\perp$ being the impact parameter Fourier conjugate to $\Delta_\perp$ represents the longitudinal component of OAM.  They may be written as $C\sim \langle s^z l^z\rangle$ where $s^z$ and $l^z$ denote the longitudinal component of the helicity and OAM of individual partons.  
Previous studies of the quark spin-orbit correlation  can be found in  \cite{Lorce:2014mxa,Rajan:2017cpx,Tan:2021osk} assuming a straight (not staple-shaped)  Wilson line, and in  \cite{Engelhardt:2021kdo} using lattice QCD. It has also been demonstrated that the quark and gluon spin-orbit correlations enter certain exclusive observables \cite{Bhattacharya:2017bvs,Bhattacharya:2018lgm,Boussarie:2018zwg,Bhattacharya:2023hbq} (denoted by $G_{1,1}$ or ${\cal G}_4$ in these references). 

In this paper, we study the small-$x$-behavior of the  $k_\perp$-moment of the 
  spin-orbit correlations defined through staple-shaped Wilson lines
\beq
C_{q,g}(x) \equiv  \int d^2k_\perp \frac{k_\perp^2}{M^2}C_{q,g} (x,k_\perp,\Delta_\perp=0). 
\label{orbit}
\eeq
 For antiquarks,  $C_{\bar{q}}(x)=-C_q(-x)$.
It is expected that $C_{q,g}(x)>0$ if the  helicity and OAM of individual quarks and gluons with a given momentum fraction $x$ tend to be aligned with each other, and $C_{q,g}(x)<0$ if they are anti-aligned. Note that this interpretation makes no reference to the spin of the parent particle. The correlation exists even in unpolarized or spinless hadrons and nuclei.

\subsection{Dipole gluon distribution}

We  demonstrate that  $C_{q,g}(x)$ at small-$x$ is calculable in the CGC framework. We begin with the dipole gluon distribution $C_g^{{\rm dip}}$.  In the limit $x\ll 1$ one can employ the eikonal approximation  $e^{i xP^+z^-}\approx 1$ and  write (\ref{a4}) as  
\beq
&& \frac{\epsilon_{ij}\int d^2w_\perp d^2z_\perp}{(2\pi)^3P^+ g^2\int d^3x } e^{-i\left(k_\perp-\frac{\Delta_\perp}{2}\right)\cdot z_\perp +i\left(k_\perp+\frac{\Delta_\perp}{2}\right)\cdot w_\perp }\langle p'|2{\rm Tr} \partial^i U(w_\perp)\partial^j U^\dagger(z_\perp)|p\rangle \nn 
&&= -\frac{N_c}{\pi \alpha_s} \epsilon_{ij}k_\perp^i\Delta_\perp^j  \frac{\int d^2w_\perp d^2 z_\perp}{(2\pi)^3}  e^{-i\left(k_\perp-\frac{\Delta_\perp}{2}\right)\cdot z_\perp +i\left(k_\perp+\frac{\Delta_\perp}{2}\right)\cdot w_\perp }
\frac{\langle p'|\frac{1}{N_c}{\rm Tr}  U(w_\perp)  U^\dagger(z_\perp)-1|p\rangle }{\langle p|p\rangle}, \label{cq}
\eeq
where  $\langle p|p\rangle =2P^+\int d^3x$ and we have integrated by parts in $z_\perp$ and $w_\perp$.  $U(w_\perp)={\rm P}\exp\left(-ig\int_{-\infty}^\infty dw^-A(w^-,w_\perp)\right)$ is an infinite Wilson line along the light-cone and their product  $\frac{1}{N_c}{\rm Tr}U(w_\perp)U^\dagger(z_\perp)$  is called the color dipole operator frequently encountered in the CGC literature. We have used the trick 
$\langle p'|\frac{1}{N_c}{\rm Tr}UU^\dagger|p\rangle = \langle p'|p\rangle + \langle  p'|\left(\frac{1}{N_c}{\rm Tr}UU^\dagger-1\right)|p\rangle$  
and dropped the first term since 
$\Delta^j_\perp \langle p'|p\rangle \propto \Delta^j_\perp \delta^{(2)}(\Delta_\perp)=0$. The forward limit $p'\to p$ can be safely taken in the second term. 
We can thus identify 
\beq
\frac{xC^{{\rm dip}}_g(x,k_\perp)}{M^2} = - \frac{2N_c}{ \alpha_s}   \int \frac{d^2w_\perp d^2 z_\perp}{(2\pi)^4}  e^{-ik_\perp\cdot (z_\perp-w_\perp)  }
\frac{\langle p|\frac{1}{N_c}{\rm Tr}  U(w_\perp)  U^\dagger(z_\perp)-1|p\rangle }{\langle p|p\rangle},
\eeq
 where the $x$-dependence is now encoded in the evolution of the dipole \cite{Balitsky:1995ub,Kovchegov:1999yj,Jalilian-Marian:1997jhx,Iancu:2000hn,Balitsky:2007feb}, see also  
\cite{Boussarie:2020fpb,Boussarie:2023xun}.
On the other hand, the unpolarized gluon TMD is obtained by simply replacing $\epsilon_{ij}\to \delta_{ij}$ and we find 
\beq
\frac{C^{{\rm dip}}_g(x,k_\perp)}{M^2}= -\frac{f^{{\rm dip}}_g(x,k_\perp)}{k_\perp^2} ,\label{gf}
\eeq
so that 
\beq
C^{{\rm dip}}_g(x) = -G(x), \label{main}
\eeq
where $G(x)$ is the unpolarized gluon PDF. The result (\ref{gf}) was previously obtained in \cite{Boer:2018vdi}, but its physical meaning was not explored. 
We see that the spin-orbit correlation does not vanish in the eikonal approximation despite its association with the polarized gluon operator $\tilde{F}^{+\mu}F^{+}_{\ \mu}$.   Moreover, it grows rapidly at small-$x$ and eventually reaches gluon saturation in exactly the same way as the unpolarized gluon distribution, but with an opposite sign. (\ref{main}) can be interpreted that the gluon helicity and OAM $s^z, l^z=\pm 1$  are `maximally' anti-aligned at small-$x$. Namely, for every single gluon at small-$x$, its  helicity and OAM are `locked-up' in opposite directions $s^z l^z=-1$. An intuitive way to understand this phenomenon is to notice that already in the leading order  production processes $q\to qg, g\to gg$, the net angular momentum of the emitted soft gluon vanishes $(s^z,l^z)=(\pm 1,\mp 1)$. The total spin-orbit correlation (\ref{main}) is therefore $-1$ times the number of gluons.

\subsection{WW gluon distribution}

Next we turn to the WW gluon distribution $C_g^{\rm WW}$. Again in the eikonal approximation, the matrix element in (\ref{a4}) can be written as  \cite{Dominguez:2011wm}, up to a prefactor,   
\beq
\epsilon_{ij}\langle p'|{\rm Tr}[U^\dagger(w_\perp) \partial_i U (w_\perp)U^\dagger(z_\perp) \partial_j U(z_\perp) ]|p\rangle . \label{mv}
\eeq  
In contrast to the dipole case, one cannot integrate by parts in $\partial_{i,j}$. As a result, a model-independent relation between $C_g^{{\rm WW}}$ and $f_g^{{\rm WW}}$ is unavailable in the present approach. We thus work in  the McLerran-Venugopalan model \cite{McLerran:1993ni,McLerran:1993ka} where the correlator of four Wilson lines (\ref{mv})  can be evaluated as \cite{Jalilian-Marian:1996mkd,McLerran:1998nk,Dominguez:2011wm} 
\beq
 C_F\epsilon_{ij}
 \frac{\partial_i^w\partial_j^z D(w_\perp,z_\perp)}{D(w_\perp,z_\perp)}\left(1- e^{\frac{N_c}{C_F}D(w_\perp,z_\perp)}\right),
 \label{cor}
\eeq  
where $C_F=\frac{N_c^2-1}{2N_c}$ and $D$ is defined through the two-point function
\beq
\frac{\langle p'|\frac{1}{N_c}{\rm Tr}U(w_\perp)U^\dagger(z_\perp)|p\rangle}{\langle p|p\rangle} = e^{D(w_\perp,z_\perp)} = e^{-i\frac{z_\perp+w_\perp}{2}\cdot \Delta_\perp}e^{D(w_\perp-z_\perp,\Delta_\perp)}.
\eeq
Changing variables as $r_\perp=z_\perp-w_\perp$ and $b_\perp=\frac{z_\perp+w_\perp}{2}$, and linearizing with respect to $\Delta_\perp$, we obtain 
\beq
k_\perp^2\frac{C^{{\rm WW}}_g(x,k_\perp)}{M^2}=-f_g^{{\rm WW}}(x,k_\perp) - \frac{C_FS_\perp}{\pi \alpha_s x} \int \frac{d^2r_\perp}{(2\pi)^3}  e^{-ik_\perp\cdot r_\perp }\partial_i^r D(r_\perp) \partial_i^r\left(\frac{  1-e^{\frac{N_c}{C_F}D(r_\perp)}  }{D(r_\perp)}\right). \label{two} 
\eeq
 where $S_\perp$ is the transverse area of the hadron and the the WW gluon TMD in the present model is well known \cite{Jalilian-Marian:1996mkd},  
\beq
f_g^{{\rm WW}}(x,k_\perp)= \frac{C_FS_\perp}{\pi\alpha_s x} \int \frac{d^2r_\perp}{(2\pi)^3}  e^{-ik_\perp\cdot r_\perp }\frac{\nabla^2 D(r_\perp)}{D(r_\perp)}\left(1- e^{\frac{N_c}{C_F}D(r_\perp)}\right). \label{fk}
\eeq  
Compare with the dipole case (\ref{gf}). The second term in (\ref{two}) is absent in the `dilute' (non-saturated) regime $D\ll 1$. We have checked numerically in a Gaussian model $D(r_\perp)\propto r_\perp^2$ that this term tends to have an opposite sign from the first term but does not flip the sign of the latter.  Moreover, the second term vanishes when integrated over $k_\perp$ due to color transparency $D(r_\perp\to 0)= 0$.  We thus find that 
\beq
C^{{\rm WW}}_g(x)= -G(x), \label{CG}
\eeq 
and therefore, $C_g^{{\rm WW}}(x)= C_g^{{\rm dip}}(x)$ at small-$x$. In fact, one can  prove that $C_g^{{\rm WW}}(x)=C_g^{{\rm dip}}(x)$ for all values of $x$,  see Appendix.   

\section{Quark spin-orbit coupling}

The analysis of the quark spin-orbit coupling is slightly more involved but basically follows the same strategy as for the quark TMD at small-$x$ \cite{McLerran:1998nk,Marquet:2009ca}.  Let us write (\ref{a3}) as
\beq
 - \frac{\int d^3z d^3w}{2(2\pi)^3\int d^3x} e^{  ixP^+(z^- -w^-)  -i\left(k_\perp-\frac{\Delta_\perp}{2}\right)\cdot z_\perp +i\left(k_\perp+\frac{\Delta_\perp}{2}\right)\cdot w_\perp} \langle p'| {\rm Tr}[W_{w,-\infty}W_{-\infty,z}\gamma^+\gamma_5S(z,w)] |p\rangle, \label{shock}
\eeq
where $S(z,w)$ is the quark propagator in the background gluon field (`shockwave') created by the high energy hadron/nucleus.  The trace is over both the color and Dirac indices. We have assumed that the staple-shaped Wilson line is past-pointing $W_-(w,z)=W_{w,-\infty}W_{-\infty,z}$. The case with a future-pointing Wilson line $W_+$ is entirely analogous. To leading order in $1/x$, one may assume that the background gauge field is localized at $z^-=0$ 
 and use the  propagator  \cite{Balitsky:1995ub}
\beq
S(z,w)= \begin{cases} \int d^4y \delta(y^-) S_0(z,y)\gamma^- S_0(y,w) U(y_\perp), \qquad (z^->0, w^-<0) \\ -\int d^4y \delta(y^-) S_0(z,y)\gamma^- S_0(y,w) U^\dagger(y_\perp), \qquad (z^-<0,w^->0) \end{cases}
\eeq
where $S_0(z,y)$ is the free fermion propagator. Note that the regions $z^-w^->0$ where $S(z,w)=S_0(z,w)$ do not contribute due to the presence of a $\gamma_5$. 
After some manipulations, (\ref{shock}) is evaluated as 
\beq 
 -\frac{iN_c}{8\pi^4x }\epsilon^{ij}\int d^2k_{g\perp} (k_\perp-k_{g\perp})_i\left[ 2\frac{k_\perp \cdot \Delta_\perp}{k_\perp^2}k_{\perp j}+ \Delta_{\perp j}  \frac{\ln  \frac{k_\perp^2}{(k_\perp-k_{g\perp})^2}}{k_\perp^2-(k_\perp-k_{g\perp})^2}\right] \frac{\langle p'|\left(\frac{1}{N_c}{\rm Tr}UU^\dagger-1\right)(k_{g\perp})|p\rangle}{\langle p|p\rangle} , \label{tilde}
\eeq
where ${\rm Tr}UU^\dagger(k_{g\perp})\equiv\int \frac{dw_\perp dz_\perp}{(2\pi)^2}e^{-ik_{g\perp}\cdot (z_\perp-w_\perp)}{\rm Tr}U(w_\perp)U^\dagger(z_\perp)$ and  we have already linearized in $\Delta_\perp$. 
 We can thus identify 
\beq
\frac{C_q(x,k_\perp)}{M^2} = \frac{N_cS_\perp}{8\pi^4x k_\perp^2} \int d^2k_{g\perp} (k_\perp-k_{g\perp})\cdot k_\perp \frac{\ln \frac{k_\perp^2}{(k_\perp-k_{g\perp})^2}}{k_\perp^2-(k_\perp-k_{g\perp})^2} \frac{\langle p|\left(\frac{1}{N_c}{\rm Tr}UU^\dagger-1\right)(k_{g\perp})|p\rangle}{\langle p|p\rangle}. \label{qg11}
\eeq
Comparing this with the known result for the quark TMD $f_q(x,k_\perp)$ \cite{McLerran:1998nk,Marquet:2009ca}, 
 we  find  the relation 
\beq
f_q (x,k_\perp)= -2\frac{k_\perp^2}{M^2}C_q(x,k_\perp), 
\eeq
or after the $k_\perp$ integration, 
\beq
C_q(x)=-\frac{1}{2}q(x), \label{quarkfin}
\eeq 
where $q(x)$ is the quark PDF. The same result holds for antiquarks $C_{\bar{q}}(x)=-\frac{1}{2}\bar{q}(x)$.  
 (\ref{quarkfin}) has a very intuitive physical explanation considering that small-$x$ quarks are dominantly created by $g\to q\bar{q}$ splittings. Suppose the parent gluon has $(s^z,l^z)=(1,-1)$ as argued above and the quark and antiquark  carry momentum fractions $1-x$ and $x$ of the gluon momentum with $x\ll 1$, respectively.  Then the quark has a higher probability to have the same helicity as the gluon, namely $s_q^z=\frac{1}{2}$, and accordingly the antiquark has the opposite helicity $s^z_{\bar{q}}=-\frac{1}{2}$  \cite{Peskin:1995ev}. Since the momentum of the quark is almost the same as that of the gluon, the OAM is the same $l^z_q=l^z_g=-1$, and  $l_{\bar{q}}^z=+1$ by overall angular momentum conservation. Therefore, in typical small-$x$ kinematics,  $(s^z,l^z)=(\pm 1,\mp 1)\to (\pm \frac{1}{2},\mp 1)\oplus (\mp\frac{1}{2},\pm 1)$. Remarkably, only the states $l^z=\pm 1$ appear in this argument although in principle $l^z$ can take any integer values.  The total quark spin-orbit correlation is thus $s^z l^z=-\frac{1}{2}$ times the number of quarks.

\section{Quantum entanglement}

We now argue that the spin-orbit anti-correlation that we have found actually signifies a deeper underlying  phenomenon; {\it quantum entanglement between helicity and OAM}. To see this, recall that the basic idea of the CGC is to  describe small-$x$ gluons as the Weisz\"acker-Williams (boosted color Coulomb) field sourced by fast-moving valence quarks at large-$x$ \cite{McLerran:1993ni}. As is well known in the context of photons in QED, such gluons are linearly polarized with  their polarization vectors randomly oriented in the transverse plane  \cite{McLerran:2001sr,Metz:2011wb}. In the helicity basis $s^z=\pm 1$, their spin states are therefore described by the  vectors 
\beq
\frac{1}{\sqrt{2}}\bigl(|+\rangle_s +|-\rangle_s\bigr), \qquad \frac{1}{\sqrt{2}i} \bigl(|+\rangle_s -|-\rangle_s\bigr), \label{ent}
\eeq
with  vanishing expectation value $\langle s^z\rangle=0$. However, we have seen above that the spin and spatial parts of the wavefunction are not independent. Given the perfect anti-correlation between helicity and  OAM $l^z=-s^z$, it is more appropriate to represent the spin states as   `Bell states'
\beq
|\Psi^+\rangle = \frac{1}{\sqrt{2}}\bigl( |+\rangle_s |-\rangle_l +|-\rangle_s |+\rangle_l\bigr), \qquad |\Psi^-\rangle =\frac{1}{\sqrt{2}i} \bigl(|+\rangle_s |-\rangle_l -|-\rangle_s |+\rangle_l \bigr),  \label{bell}
\eeq
such that  $\langle \Psi^\pm|s^z l^z|\Psi^\pm\rangle = -1$ even though $\langle \Psi^\pm|s^z |\Psi^\pm\rangle=0$ and $\langle \Psi^\pm|l^z |\Psi^\pm\rangle=0$. These states are also linearly polarized (since $\langle s^z\rangle=0$) with the orientation of the polarization vector varying spatially.  For example, the so-called vortex mode of photons carries OAM characterized by the azimuthal angular wavefunction $|\pm \rangle_l \sim e^{\pm i\phi}$ \cite{Bliokh:2015doa}. This leads to angular-dependent polarization vectors  $|\Psi^+\rangle \sim (1,i) e^{-i\phi} + (1,-i)e^{i\phi} \sim (\cos \phi,\sin\phi)$ 
and $|\Psi^-\rangle \sim -i\left((1,i)e^{-i\phi} -(1,-i)e^{i\phi}\right) \sim (-\sin \phi, \cos \phi)$. 
 In QCD, the OAM of  gluons is associated with their literal orbital motion inside a confined hadron. The corresponding wavefunction will be more complicated.

$|\Psi^+\rangle$ and $|\Psi^-\rangle$ are maximally entangled states realized on a  single gluon at small-$x$ each carrying  entanglement entropy $\ln 2$. 
Remarkably, a quite analogous phenomenon, quantum entanglement of the spin and the OAM of a single photon, has been experimentally realized in the field of quantum optics \cite{stav}. In this experiment, the transition from (\ref{ent}) to (\ref{bell}) was artificially made by passing linearly polarized photons  through a specially engineered surface which creates or annihilates one unit of OAM depending on the photon helicity.  In contrast, in QCD the spin-orbit entanglement is a default property of soft gluons from birth. 

 Note that some kind of entanglement exists for all values of $x$ as long as $C_{q,g}(x)\neq 0$. The previous results regarding the integral $C_q=\int_0^1dx C_q(x)<0$ \cite{Lorce:2014mxa,Tan:2021osk,Engelhardt:2021kdo,Kim:2024cbq} may be reinterpreted from this perspective. However, for generic values of $x$, there is no constraint on the allowed values of OAM  $l^z=0,\pm 1, \pm 2,\cdots$.   It it is only in the small-$x$ region where only  the states $l^z=\pm 1$ are selected by kinematics and these states form a  `qubit'. This is  perfectly  anti-correlated with another qubit in the spin space $s^z=\pm 1$ to form  the maximally entangled Bell states (\ref{bell}). Thus the phenomenon is unique to small-$x$ and represents  a novel form of entanglement  distinct from those previously discussed  in the small-$x$ literature  \cite{Kutak:2011rb,Kovner:2015hga,Kharzeev:2017qzs,Hagiwara:2017uaz,Neill:2018uqw,Dvali:2021ooc,Dumitru:2023qee}. 

\section{Conclusions} 

In this paper, we have uncovered a novel emergent  property of the Color Glass Condensate at high energy. We have shown  that the  helicity  and OAM of single quarks and gluons at small-$x$ are strongly anti-aligned  even if the parent hadron/nucleus is unpolarized or spinless. 
Combining the results with the  fact that soft gluons in the CGC are linearly polarized, we conclude that the helicity and OAM of soft gluons are maximally entangled in a quantum mechanical sense. In the parlance of  quantum information science,  neither the measurement of helicity $\langle s^z\rangle =0$ nor OAM $\langle l^z\rangle=0$ alone can reveal the true quantum nature of the state which is encoded in the correlation $\langle s^zl^z\rangle\neq 0$.

Our finding is consistent with and even explains the known anti-correlation between the quark $\Delta q(x)$ and gluon $\Delta G(x)$ helicity  distributions and the OAM $L_{q,g}(x)$ distributions in a {\it polarized} hadron \cite{Hatta:2016aoc,More:2017zqp,Hatta:2018itc,Boussarie:2019icw,Kovchegov:2023yzd,Manley:2024pcl}. 
When a hadron is polarized, the helicities of quarks and gluons tend to align in certain directions. Since the helicity and OAM of each  parton are locked up even before polarization, naturally the resulting  distributions have opposite signs at small-$x$ \cite{Hatta:2018itc,Boussarie:2019icw}
\beq
\Delta q(x) \approx -\frac{1}{1+c}L_q(x), \qquad \Delta G(x)\approx -\frac{2}{1+c}L_g(x), \label{22}
\eeq
where $c$ is the Regge intercept $\Delta q(x)\sim \Delta G(x) \sim 1/x^c$. Note however that perfect anti-correlation would imply $\Delta q(x)\approx -L_q(x)$ and $\Delta G(x) \approx -L_g(x)$, but this does not agree with (\ref{22}) for any value of $c$  (see also \cite{Kovchegov:2023yzd,Manley:2024pcl}). 
It would be interesting to further investigate the connection between these observations. 

Finally, it is of course very interesting to experimentally probe the spin-orbit correlation. Previously the quark and gluon $G_{1,1}$ GTMDs have appeared in certain exclusive reactions   
\cite{Bhattacharya:2017bvs,Bhattacharya:2018lgm,Boussarie:2018zwg}, but a quantitative evaluation of cross sections in a concrete experimental setup has been lacking. We have recently found that the gluon $G_{1,1}$ at small-$x$  affects the observable  proposed in \cite{Bhattacharya:2022vvo}. This will be elaborated in a forthcoming publication  \cite{Bhattacharya:2024sck}

\section*{Acknowledgements}

The work of S.~B. has been supported by the Laboratory Directed Research and Development program of Los Alamos National Laboratory under project number 20240738PRD1.
S.~B. has also received support from the U.~S. Department of Energy through the Los Alamos National Laboratory. Los Alamos National Laboratory is operated by Triad National Security, LLC, for the National Nuclear Security Administration of U.~S. Department of Energy (Contract No. 89233218CNA000001).
Y.~H. was supported by the U.S. Department of Energy under Contract No. DE-SC0012704, and also by  Laboratory Directed Research and Development (LDRD) funds from Brookhaven Science Associates. R.~B. and Y.~H. were also supported  by the framework of the Saturated Glue (SURGE)
Topical Theory Collaboration.

\section*{Appendix}

Here we prove that $C_g^{\rm WW}(x)=C_g^{\rm dip}(x)$ for all values of $x$. Our proof is similar to the related proof in \cite{Hatta:2016aoc} that the gluon orbital angular momentum distributions defined from the WW and dipole gluon distributions are identical $L_g^{\rm WW}(x)=L_g^{\rm dip}(x)$.  Following the derivation in  Section III of  \cite{Hatta:2016aoc} (see (36) there), we find that the difference between $C_g^{\rm WW}(x)$ and $C_g^{\rm dip}(x)$ boils down to the property of the  matrix element 
\beq
\int dz^- e^{ixP^+z^-}\langle p'|\int_{-\infty}^\infty dy^- {\rm Tr}[\tilde{F}^{+i}(-z^-/2)F^{+j}(y^-)F^{+i}(z^-/2)]|p\rangle, \label{fff}
\eeq
where the three fields are collinear along the light-like direction and Wilson lines in between are understood. If (\ref{fff}) has a matrix element proportional to  $\epsilon^{jl}\Delta_l$, this contributes to the difference $C_g^{\rm WW}(x)-C_g^{\rm dip}(x)$. To see this is forbidden, consider the $PT$ transform of (\ref{fff}). 
\beq
\langle p'|\int dy^- {\rm Tr}[\tilde{F}^{+i}(-z^-/2)F^{+j}(y^-)F^{+i}(z^-/2)]|p\rangle  &=& -\langle p|\int dy^- {\rm Tr}[F^{+i}(-z^-/2)F^{+j}(-y^-)\tilde{F}^{+i}(z^-/2)]|p'\rangle \nn 
&=&  \langle p|\int dy^- {\rm Tr}[\tilde{F}^{+i}(-z^-/2)F^{+j}(y^-)F^{+i}(z^-/2)]|p'\rangle. \label{ffff}
\eeq
Since $\Delta \to -\Delta$ under $p\leftrightarrow p'$, the linear term $\epsilon^{jl}\Delta_l$ cannot arise and thus $C_g^{\rm WW}(x)=C_q^{\rm dip}(x)$.   Note that in \cite{Hatta:2016aoc}, the nucleon spin vector provides a minus sign $S^+\to -S^+$ under $PT$, whereas in (\ref{ffff}) the sign flip in the second line is due to the antisymmetric tensor $F^{+i}\cdots\tilde{F}^{+i}=-\tilde{F}^{+i}\cdots F^{+i}$.

\bibliography{ref}

\begin{thebibliography}{55}%
\makeatletter
\providecommand \@ifxundefined [1]{%
 \@ifx{#1\undefined}
}%
\providecommand \@ifnum [1]{%
 \ifnum #1\expandafter \@firstoftwo
 \else \expandafter \@secondoftwo
 \fi
}%
\providecommand \@ifx [1]{%
 \ifx #1\expandafter \@firstoftwo
 \else \expandafter \@secondoftwo
 \fi
}%
\providecommand \natexlab [1]{#1}%
\providecommand \enquote  [1]{``#1''}%
\providecommand \bibnamefont  [1]{#1}%
\providecommand \bibfnamefont [1]{#1}%
\providecommand \citenamefont [1]{#1}%
\providecommand \href@noop [0]{\@secondoftwo}%
\providecommand \href [0]{\begingroup \@sanitize@url \@href}%
\providecommand \@href[1]{\@@startlink{#1}\@@href}%
\providecommand \@@href[1]{\endgroup#1\@@endlink}%
\providecommand \@sanitize@url [0]{\catcode `\\12\catcode `\$12\catcode `\&12\catcode `\#12\catcode `\^12\catcode `\_12\catcode `\%12\relax}%
\providecommand \@@startlink[1]{}%
\providecommand \@@endlink[0]{}%
\providecommand \url  [0]{\begingroup\@sanitize@url \@url }%
\providecommand \@url [1]{\endgroup\@href {#1}{\urlprefix }}%
\providecommand \urlprefix  [0]{URL }%
\providecommand \Eprint [0]{\href }%
\providecommand \doibase [0]{http://dx.doi.org/}%
\providecommand \selectlanguage [0]{\@gobble}%
\providecommand \bibinfo  [0]{\@secondoftwo}%
\providecommand \bibfield  [0]{\@secondoftwo}%
\providecommand \translation [1]{[#1]}%
\providecommand \BibitemOpen [0]{}%
\providecommand \bibitemStop [0]{}%
\providecommand \bibitemNoStop [0]{.\EOS\space}%
\providecommand \EOS [0]{\spacefactor3000\relax}%
\providecommand \BibitemShut  [1]{\csname bibitem#1\endcsname}%
\let\auto@bib@innerbib\@empty
\bibitem [{\citenamefont {Abdul~Khalek}\ \emph {et~al.}(2022)\citenamefont {Abdul~Khalek} \emph {et~al.}}]{AbdulKhalek:2021gbh}%
  \BibitemOpen
  \bibfield  {author} {\bibinfo {author} {\bibfnamefont {R.}~\bibnamefont {Abdul~Khalek}} \emph {et~al.},\ }\href {\doibase 10.1016/j.nuclphysa.2022.122447} {\bibfield  {journal} {\bibinfo  {journal} {Nucl. Phys. A}\ }\textbf {\bibinfo {volume} {1026}},\ \bibinfo {pages} {122447} (\bibinfo {year} {2022})},\ \Eprint {http://arxiv.org/abs/2103.05419} {arXiv:2103.05419 [physics.ins-det]} \BibitemShut {NoStop}%
\bibitem [{\citenamefont {Gribov}\ \emph {et~al.}(1983)\citenamefont {Gribov}, \citenamefont {Levin},\ and\ \citenamefont {Ryskin}}]{Gribov:1983ivg}%
  \BibitemOpen
  \bibfield  {author} {\bibinfo {author} {\bibfnamefont {L.~V.}\ \bibnamefont {Gribov}}, \bibinfo {author} {\bibfnamefont {E.~M.}\ \bibnamefont {Levin}}, \ and\ \bibinfo {author} {\bibfnamefont {M.~G.}\ \bibnamefont {Ryskin}},\ }\href {\doibase 10.1016/0370-1573(83)90022-4} {\bibfield  {journal} {\bibinfo  {journal} {Phys. Rept.}\ }\textbf {\bibinfo {volume} {100}},\ \bibinfo {pages} {1} (\bibinfo {year} {1983})}\BibitemShut {NoStop}%
\bibitem [{\citenamefont {Gelis}\ \emph {et~al.}(2010)\citenamefont {Gelis}, \citenamefont {Iancu}, \citenamefont {Jalilian-Marian},\ and\ \citenamefont {Venugopalan}}]{Gelis:2010nm}%
  \BibitemOpen
  \bibfield  {author} {\bibinfo {author} {\bibfnamefont {F.}~\bibnamefont {Gelis}}, \bibinfo {author} {\bibfnamefont {E.}~\bibnamefont {Iancu}}, \bibinfo {author} {\bibfnamefont {J.}~\bibnamefont {Jalilian-Marian}}, \ and\ \bibinfo {author} {\bibfnamefont {R.}~\bibnamefont {Venugopalan}},\ }\href {\doibase 10.1146/annurev.nucl.010909.083629} {\bibfield  {journal} {\bibinfo  {journal} {Ann. Rev. Nucl. Part. Sci.}\ }\textbf {\bibinfo {volume} {60}},\ \bibinfo {pages} {463} (\bibinfo {year} {2010})},\ \Eprint {http://arxiv.org/abs/1002.0333} {arXiv:1002.0333 [hep-ph]} \BibitemShut {NoStop}%
\bibitem [{\citenamefont {Morreale}\ and\ \citenamefont {Salazar}(2021)}]{Morreale:2021pnn}%
  \BibitemOpen
  \bibfield  {author} {\bibinfo {author} {\bibfnamefont {A.}~\bibnamefont {Morreale}}\ and\ \bibinfo {author} {\bibfnamefont {F.}~\bibnamefont {Salazar}},\ }\href {\doibase 10.3390/universe7080312} {\bibfield  {journal} {\bibinfo  {journal} {Universe}\ }\textbf {\bibinfo {volume} {7}},\ \bibinfo {pages} {312} (\bibinfo {year} {2021})},\ \Eprint {http://arxiv.org/abs/2108.08254} {arXiv:2108.08254 [hep-ph]} \BibitemShut {NoStop}%
\bibitem [{\citenamefont {{National Academies of Sciences, Engineering, and Medicine 2018}}()}]{nas}%
  \BibitemOpen
  \bibfield  {author} {\bibinfo {author} {\bibnamefont {{National Academies of Sciences, Engineering, and Medicine 2018}}},\ }\href@noop {} {\enquote {\bibinfo {title} {{An Assessment of U.S.-Based Electron-Ion Collider Science}},}\ }\bibinfo {howpublished} {\url{https://doi.org/10.17226/25171}}\BibitemShut {NoStop}%
\bibitem [{\citenamefont {Lorce}\ and\ \citenamefont {Pasquini}(2011)}]{Lorce:2011kd}%
  \BibitemOpen
  \bibfield  {author} {\bibinfo {author} {\bibfnamefont {C.}~\bibnamefont {Lorce}}\ and\ \bibinfo {author} {\bibfnamefont {B.}~\bibnamefont {Pasquini}},\ }\href {\doibase 10.1103/PhysRevD.84.014015} {\bibfield  {journal} {\bibinfo  {journal} {Phys. Rev. D}\ }\textbf {\bibinfo {volume} {84}},\ \bibinfo {pages} {014015} (\bibinfo {year} {2011})},\ \Eprint {http://arxiv.org/abs/1106.0139} {arXiv:1106.0139 [hep-ph]} \BibitemShut {NoStop}%
\bibitem [{\citenamefont {Lorc\'e}(2014)}]{Lorce:2014mxa}%
  \BibitemOpen
  \bibfield  {author} {\bibinfo {author} {\bibfnamefont {C.}~\bibnamefont {Lorc\'e}},\ }\href {\doibase 10.1016/j.physletb.2014.06.068} {\bibfield  {journal} {\bibinfo  {journal} {Phys. Lett. B}\ }\textbf {\bibinfo {volume} {735}},\ \bibinfo {pages} {344} (\bibinfo {year} {2014})},\ \Eprint {http://arxiv.org/abs/1401.7784} {arXiv:1401.7784 [hep-ph]} \BibitemShut {NoStop}%
\bibitem [{\citenamefont {Rajan}\ \emph {et~al.}(2018)\citenamefont {Rajan}, \citenamefont {Engelhardt},\ and\ \citenamefont {Liuti}}]{Rajan:2017cpx}%
  \BibitemOpen
  \bibfield  {author} {\bibinfo {author} {\bibfnamefont {A.}~\bibnamefont {Rajan}}, \bibinfo {author} {\bibfnamefont {M.}~\bibnamefont {Engelhardt}}, \ and\ \bibinfo {author} {\bibfnamefont {S.}~\bibnamefont {Liuti}},\ }\href {\doibase 10.1103/PhysRevD.98.074022} {\bibfield  {journal} {\bibinfo  {journal} {Phys. Rev. D}\ }\textbf {\bibinfo {volume} {98}},\ \bibinfo {pages} {074022} (\bibinfo {year} {2018})},\ \Eprint {http://arxiv.org/abs/1709.05770} {arXiv:1709.05770 [hep-ph]} \BibitemShut {NoStop}%
\bibitem [{\citenamefont {Tan}\ and\ \citenamefont {Lu}(2022)}]{Tan:2021osk}%
  \BibitemOpen
  \bibfield  {author} {\bibinfo {author} {\bibfnamefont {C.}~\bibnamefont {Tan}}\ and\ \bibinfo {author} {\bibfnamefont {Z.}~\bibnamefont {Lu}},\ }\href {\doibase 10.1103/PhysRevD.105.034004} {\bibfield  {journal} {\bibinfo  {journal} {Phys. Rev. D}\ }\textbf {\bibinfo {volume} {105}},\ \bibinfo {pages} {034004} (\bibinfo {year} {2022})},\ \Eprint {http://arxiv.org/abs/2110.08502} {arXiv:2110.08502 [hep-ph]} \BibitemShut {NoStop}%
\bibitem [{\citenamefont {Engelhardt}\ \emph {et~al.}(2022)\citenamefont {Engelhardt} \emph {et~al.}}]{Engelhardt:2021kdo}%
  \BibitemOpen
  \bibfield  {author} {\bibinfo {author} {\bibfnamefont {M.}~\bibnamefont {Engelhardt}} \emph {et~al.},\ }\href {\doibase 10.22323/1.396.0413} {\bibfield  {journal} {\bibinfo  {journal} {PoS}\ }\textbf {\bibinfo {volume} {LATTICE2021}},\ \bibinfo {pages} {413} (\bibinfo {year} {2022})},\ \Eprint {http://arxiv.org/abs/2112.13464} {arXiv:2112.13464 [hep-lat]} \BibitemShut {NoStop}%
\bibitem [{\citenamefont {Kim}\ \emph {et~al.}(2024)\citenamefont {Kim}, \citenamefont {Won}, \citenamefont {Kim},\ and\ \citenamefont {Weiss}}]{Kim:2024cbq}%
  \BibitemOpen
  \bibfield  {author} {\bibinfo {author} {\bibfnamefont {J.-Y.}\ \bibnamefont {Kim}}, \bibinfo {author} {\bibfnamefont {H.-Y.}\ \bibnamefont {Won}}, \bibinfo {author} {\bibfnamefont {H.-C.}\ \bibnamefont {Kim}}, \ and\ \bibinfo {author} {\bibfnamefont {C.}~\bibnamefont {Weiss}},\ }\href@noop {} {\  (\bibinfo {year} {2024})},\ \Eprint {http://arxiv.org/abs/2403.07186} {arXiv:2403.07186 [hep-ph]} \BibitemShut {NoStop}%
\bibitem [{\citenamefont {Bhattacharya}\ \emph {et~al.}(2017)\citenamefont {Bhattacharya}, \citenamefont {Metz},\ and\ \citenamefont {Zhou}}]{Bhattacharya:2017bvs}%
  \BibitemOpen
  \bibfield  {author} {\bibinfo {author} {\bibfnamefont {S.}~\bibnamefont {Bhattacharya}}, \bibinfo {author} {\bibfnamefont {A.}~\bibnamefont {Metz}}, \ and\ \bibinfo {author} {\bibfnamefont {J.}~\bibnamefont {Zhou}},\ }\href {\doibase 10.1016/j.physletb.2017.05.081} {\bibfield  {journal} {\bibinfo  {journal} {Phys. Lett. B}\ }\textbf {\bibinfo {volume} {771}},\ \bibinfo {pages} {396} (\bibinfo {year} {2017})},\ \bibinfo {note} {[Erratum: Phys.Lett.B 810, 135866 (2020)]},\ \Eprint {http://arxiv.org/abs/1702.04387} {arXiv:1702.04387 [hep-ph]} \BibitemShut {NoStop}%
\bibitem [{\citenamefont {Bhattacharya}\ \emph {et~al.}(2022{\natexlab{a}})\citenamefont {Bhattacharya}, \citenamefont {Metz}, \citenamefont {Ojha}, \citenamefont {Tsai},\ and\ \citenamefont {Zhou}}]{Bhattacharya:2018lgm}%
  \BibitemOpen
  \bibfield  {author} {\bibinfo {author} {\bibfnamefont {S.}~\bibnamefont {Bhattacharya}}, \bibinfo {author} {\bibfnamefont {A.}~\bibnamefont {Metz}}, \bibinfo {author} {\bibfnamefont {V.~K.}\ \bibnamefont {Ojha}}, \bibinfo {author} {\bibfnamefont {J.-Y.}\ \bibnamefont {Tsai}}, \ and\ \bibinfo {author} {\bibfnamefont {J.}~\bibnamefont {Zhou}},\ }\href {\doibase 10.1016/j.physletb.2022.137383} {\bibfield  {journal} {\bibinfo  {journal} {Phys. Lett. B}\ }\textbf {\bibinfo {volume} {833}},\ \bibinfo {pages} {137383} (\bibinfo {year} {2022}{\natexlab{a}})},\ \Eprint {http://arxiv.org/abs/1802.10550} {arXiv:1802.10550 [hep-ph]} \BibitemShut {NoStop}%
\bibitem [{\citenamefont {Boussarie}\ \emph {et~al.}(2018)\citenamefont {Boussarie}, \citenamefont {Hatta}, \citenamefont {Xiao},\ and\ \citenamefont {Yuan}}]{Boussarie:2018zwg}%
  \BibitemOpen
  \bibfield  {author} {\bibinfo {author} {\bibfnamefont {R.}~\bibnamefont {Boussarie}}, \bibinfo {author} {\bibfnamefont {Y.}~\bibnamefont {Hatta}}, \bibinfo {author} {\bibfnamefont {B.-W.}\ \bibnamefont {Xiao}}, \ and\ \bibinfo {author} {\bibfnamefont {F.}~\bibnamefont {Yuan}},\ }\href {\doibase 10.1103/PhysRevD.98.074015} {\bibfield  {journal} {\bibinfo  {journal} {Phys. Rev. D}\ }\textbf {\bibinfo {volume} {98}},\ \bibinfo {pages} {074015} (\bibinfo {year} {2018})},\ \Eprint {http://arxiv.org/abs/1807.08697} {arXiv:1807.08697 [hep-ph]} \BibitemShut {NoStop}%
\bibitem [{\citenamefont {Bhattacharya}\ \emph {et~al.}(2023)\citenamefont {Bhattacharya}, \citenamefont {Zheng},\ and\ \citenamefont {Zhou}}]{Bhattacharya:2023hbq}%
  \BibitemOpen
  \bibfield  {author} {\bibinfo {author} {\bibfnamefont {S.}~\bibnamefont {Bhattacharya}}, \bibinfo {author} {\bibfnamefont {D.}~\bibnamefont {Zheng}}, \ and\ \bibinfo {author} {\bibfnamefont {J.}~\bibnamefont {Zhou}},\ }\href@noop {} {\  (\bibinfo {year} {2023})},\ \Eprint {http://arxiv.org/abs/2312.01309} {arXiv:2312.01309 [hep-ph]} \BibitemShut {NoStop}%
\bibitem [{\citenamefont {Boer}\ \emph {et~al.}(2018)\citenamefont {Boer}, \citenamefont {Van~Daal}, \citenamefont {Mulders},\ and\ \citenamefont {Petreska}}]{Boer:2018vdi}%
  \BibitemOpen
  \bibfield  {author} {\bibinfo {author} {\bibfnamefont {D.}~\bibnamefont {Boer}}, \bibinfo {author} {\bibfnamefont {T.}~\bibnamefont {Van~Daal}}, \bibinfo {author} {\bibfnamefont {P.~J.}\ \bibnamefont {Mulders}}, \ and\ \bibinfo {author} {\bibfnamefont {E.}~\bibnamefont {Petreska}},\ }\href {\doibase 10.1007/JHEP07(2018)140} {\bibfield  {journal} {\bibinfo  {journal} {JHEP}\ }\textbf {\bibinfo {volume} {07}},\ \bibinfo {pages} {140} (\bibinfo {year} {2018})},\ \Eprint {http://arxiv.org/abs/1805.05219} {arXiv:1805.05219 [hep-ph]} \BibitemShut {NoStop}%
\bibitem [{\citenamefont {Kuraev}\ \emph {et~al.}(1977)\citenamefont {Kuraev}, \citenamefont {Lipatov},\ and\ \citenamefont {Fadin}}]{Kuraev:1977fs}%
  \BibitemOpen
  \bibfield  {author} {\bibinfo {author} {\bibfnamefont {E.~A.}\ \bibnamefont {Kuraev}}, \bibinfo {author} {\bibfnamefont {L.~N.}\ \bibnamefont {Lipatov}}, \ and\ \bibinfo {author} {\bibfnamefont {V.~S.}\ \bibnamefont {Fadin}},\ }\href@noop {} {\bibfield  {journal} {\bibinfo  {journal} {Sov. Phys. JETP}\ }\textbf {\bibinfo {volume} {45}},\ \bibinfo {pages} {199} (\bibinfo {year} {1977})}\BibitemShut {NoStop}%
\bibitem [{\citenamefont {Balitsky}\ and\ \citenamefont {Lipatov}(1978)}]{Balitsky:1978ic}%
  \BibitemOpen
  \bibfield  {author} {\bibinfo {author} {\bibfnamefont {I.~I.}\ \bibnamefont {Balitsky}}\ and\ \bibinfo {author} {\bibfnamefont {L.~N.}\ \bibnamefont {Lipatov}},\ }\href@noop {} {\bibfield  {journal} {\bibinfo  {journal} {Sov. J. Nucl. Phys.}\ }\textbf {\bibinfo {volume} {28}},\ \bibinfo {pages} {822} (\bibinfo {year} {1978})}\BibitemShut {NoStop}%
\bibitem [{\citenamefont {Cougoulic}\ \emph {et~al.}(2022)\citenamefont {Cougoulic}, \citenamefont {Kovchegov}, \citenamefont {Tarasov},\ and\ \citenamefont {Tawabutr}}]{Cougoulic:2022gbk}%
  \BibitemOpen
  \bibfield  {author} {\bibinfo {author} {\bibfnamefont {F.}~\bibnamefont {Cougoulic}}, \bibinfo {author} {\bibfnamefont {Y.~V.}\ \bibnamefont {Kovchegov}}, \bibinfo {author} {\bibfnamefont {A.}~\bibnamefont {Tarasov}}, \ and\ \bibinfo {author} {\bibfnamefont {Y.}~\bibnamefont {Tawabutr}},\ }\href {\doibase 10.1007/JHEP07(2022)095} {\bibfield  {journal} {\bibinfo  {journal} {JHEP}\ }\textbf {\bibinfo {volume} {07}},\ \bibinfo {pages} {095} (\bibinfo {year} {2022})},\ \Eprint {http://arxiv.org/abs/2204.11898} {arXiv:2204.11898 [hep-ph]} \BibitemShut {NoStop}%
\bibitem [{\citenamefont {Hatta}\ \emph {et~al.}(2017)\citenamefont {Hatta}, \citenamefont {Nakagawa}, \citenamefont {Yuan}, \citenamefont {Zhao},\ and\ \citenamefont {Xiao}}]{Hatta:2016aoc}%
  \BibitemOpen
  \bibfield  {author} {\bibinfo {author} {\bibfnamefont {Y.}~\bibnamefont {Hatta}}, \bibinfo {author} {\bibfnamefont {Y.}~\bibnamefont {Nakagawa}}, \bibinfo {author} {\bibfnamefont {F.}~\bibnamefont {Yuan}}, \bibinfo {author} {\bibfnamefont {Y.}~\bibnamefont {Zhao}}, \ and\ \bibinfo {author} {\bibfnamefont {B.}~\bibnamefont {Xiao}},\ }\href {\doibase 10.1103/PhysRevD.95.114032} {\bibfield  {journal} {\bibinfo  {journal} {Phys. Rev. D}\ }\textbf {\bibinfo {volume} {95}},\ \bibinfo {pages} {114032} (\bibinfo {year} {2017})},\ \Eprint {http://arxiv.org/abs/1612.02445} {arXiv:1612.02445 [hep-ph]} \BibitemShut {NoStop}%
\bibitem [{\citenamefont {More}\ \emph {et~al.}(2018)\citenamefont {More}, \citenamefont {Mukherjee},\ and\ \citenamefont {Nair}}]{More:2017zqp}%
  \BibitemOpen
  \bibfield  {author} {\bibinfo {author} {\bibfnamefont {J.}~\bibnamefont {More}}, \bibinfo {author} {\bibfnamefont {A.}~\bibnamefont {Mukherjee}}, \ and\ \bibinfo {author} {\bibfnamefont {S.}~\bibnamefont {Nair}},\ }\href {\doibase 10.1140/epjc/s10052-018-5858-1} {\bibfield  {journal} {\bibinfo  {journal} {Eur. Phys. J. C}\ }\textbf {\bibinfo {volume} {78}},\ \bibinfo {pages} {389} (\bibinfo {year} {2018})},\ \Eprint {http://arxiv.org/abs/1709.00943} {arXiv:1709.00943 [hep-ph]} \BibitemShut {NoStop}%
\bibitem [{\citenamefont {Hatta}\ and\ \citenamefont {Yang}(2018)}]{Hatta:2018itc}%
  \BibitemOpen
  \bibfield  {author} {\bibinfo {author} {\bibfnamefont {Y.}~\bibnamefont {Hatta}}\ and\ \bibinfo {author} {\bibfnamefont {D.-J.}\ \bibnamefont {Yang}},\ }\href {\doibase 10.1016/j.physletb.2018.03.081} {\bibfield  {journal} {\bibinfo  {journal} {Phys. Lett. B}\ }\textbf {\bibinfo {volume} {781}},\ \bibinfo {pages} {213} (\bibinfo {year} {2018})},\ \Eprint {http://arxiv.org/abs/1802.02716} {arXiv:1802.02716 [hep-ph]} \BibitemShut {NoStop}%
\bibitem [{\citenamefont {Boussarie}\ \emph {et~al.}(2019)\citenamefont {Boussarie}, \citenamefont {Hatta},\ and\ \citenamefont {Yuan}}]{Boussarie:2019icw}%
  \BibitemOpen
  \bibfield  {author} {\bibinfo {author} {\bibfnamefont {R.}~\bibnamefont {Boussarie}}, \bibinfo {author} {\bibfnamefont {Y.}~\bibnamefont {Hatta}}, \ and\ \bibinfo {author} {\bibfnamefont {F.}~\bibnamefont {Yuan}},\ }\href {\doibase 10.1016/j.physletb.2019.134817} {\bibfield  {journal} {\bibinfo  {journal} {Phys. Lett. B}\ }\textbf {\bibinfo {volume} {797}},\ \bibinfo {pages} {134817} (\bibinfo {year} {2019})},\ \Eprint {http://arxiv.org/abs/1904.02693} {arXiv:1904.02693 [hep-ph]} \BibitemShut {NoStop}%
\bibitem [{\citenamefont {Kovchegov}\ and\ \citenamefont {Manley}(2024)}]{Kovchegov:2023yzd}%
  \BibitemOpen
  \bibfield  {author} {\bibinfo {author} {\bibfnamefont {Y.~V.}\ \bibnamefont {Kovchegov}}\ and\ \bibinfo {author} {\bibfnamefont {B.}~\bibnamefont {Manley}},\ }\href {\doibase 10.1007/JHEP02(2024)060} {\bibfield  {journal} {\bibinfo  {journal} {JHEP}\ }\textbf {\bibinfo {volume} {02}},\ \bibinfo {pages} {060} (\bibinfo {year} {2024})},\ \Eprint {http://arxiv.org/abs/2310.18404} {arXiv:2310.18404 [hep-ph]} \BibitemShut {NoStop}%
\bibitem [{\citenamefont {Manley}(2024)}]{Manley:2024pcl}%
  \BibitemOpen
  \bibfield  {author} {\bibinfo {author} {\bibfnamefont {B.}~\bibnamefont {Manley}},\ }\href@noop {} {\  (\bibinfo {year} {2024})},\ \Eprint {http://arxiv.org/abs/2401.05508} {arXiv:2401.05508 [hep-ph]} \BibitemShut {NoStop}%
\bibitem [{\citenamefont {Stav}\ \emph {et~al.}(2018)\citenamefont {Stav}, \citenamefont {Faerman}, \citenamefont {Maguid}, \citenamefont {Oren}, \citenamefont {Kleiner}, \citenamefont {Hasman},\ and\ \citenamefont {Segev}}]{stav}%
  \BibitemOpen
  \bibfield  {author} {\bibinfo {author} {\bibfnamefont {T.}~\bibnamefont {Stav}}, \bibinfo {author} {\bibfnamefont {A.}~\bibnamefont {Faerman}}, \bibinfo {author} {\bibfnamefont {E.}~\bibnamefont {Maguid}}, \bibinfo {author} {\bibfnamefont {D.}~\bibnamefont {Oren}}, \bibinfo {author} {\bibfnamefont {V.}~\bibnamefont {Kleiner}}, \bibinfo {author} {\bibfnamefont {E.}~\bibnamefont {Hasman}}, \ and\ \bibinfo {author} {\bibfnamefont {M.}~\bibnamefont {Segev}},\ }\href@noop {} {\bibfield  {journal} {\bibinfo  {journal} {Science}\ }\textbf {\bibinfo {volume} {361}},\ \bibinfo {pages} {1101} (\bibinfo {year} {2018})},\ \Eprint {http://arxiv.org/abs/1802.06374} {arXiv:1802.06374} \BibitemShut {NoStop}%
\bibitem [{\citenamefont {Bomhof}\ \emph {et~al.}(2006)\citenamefont {Bomhof}, \citenamefont {Mulders},\ and\ \citenamefont {Pijlman}}]{Bomhof:2006dp}%
  \BibitemOpen
  \bibfield  {author} {\bibinfo {author} {\bibfnamefont {C.~J.}\ \bibnamefont {Bomhof}}, \bibinfo {author} {\bibfnamefont {P.~J.}\ \bibnamefont {Mulders}}, \ and\ \bibinfo {author} {\bibfnamefont {F.}~\bibnamefont {Pijlman}},\ }\href {\doibase 10.1140/epjc/s2006-02554-2} {\bibfield  {journal} {\bibinfo  {journal} {Eur. Phys. J. C}\ }\textbf {\bibinfo {volume} {47}},\ \bibinfo {pages} {147} (\bibinfo {year} {2006})},\ \Eprint {http://arxiv.org/abs/hep-ph/0601171} {arXiv:hep-ph/0601171} \BibitemShut {NoStop}%
\bibitem [{\citenamefont {Dominguez}\ \emph {et~al.}(2011)\citenamefont {Dominguez}, \citenamefont {Marquet}, \citenamefont {Xiao},\ and\ \citenamefont {Yuan}}]{Dominguez:2011wm}%
  \BibitemOpen
  \bibfield  {author} {\bibinfo {author} {\bibfnamefont {F.}~\bibnamefont {Dominguez}}, \bibinfo {author} {\bibfnamefont {C.}~\bibnamefont {Marquet}}, \bibinfo {author} {\bibfnamefont {B.-W.}\ \bibnamefont {Xiao}}, \ and\ \bibinfo {author} {\bibfnamefont {F.}~\bibnamefont {Yuan}},\ }\href {\doibase 10.1103/PhysRevD.83.105005} {\bibfield  {journal} {\bibinfo  {journal} {Phys. Rev. D}\ }\textbf {\bibinfo {volume} {83}},\ \bibinfo {pages} {105005} (\bibinfo {year} {2011})},\ \Eprint {http://arxiv.org/abs/1101.0715} {arXiv:1101.0715 [hep-ph]} \BibitemShut {NoStop}%
\bibitem [{\citenamefont {Meissner}\ \emph {et~al.}(2008)\citenamefont {Meissner}, \citenamefont {Metz}, \citenamefont {Schlegel},\ and\ \citenamefont {Goeke}}]{Meissner:2008ay}%
  \BibitemOpen
  \bibfield  {author} {\bibinfo {author} {\bibfnamefont {S.}~\bibnamefont {Meissner}}, \bibinfo {author} {\bibfnamefont {A.}~\bibnamefont {Metz}}, \bibinfo {author} {\bibfnamefont {M.}~\bibnamefont {Schlegel}}, \ and\ \bibinfo {author} {\bibfnamefont {K.}~\bibnamefont {Goeke}},\ }\href {\doibase 10.1088/1126-6708/2008/08/038} {\bibfield  {journal} {\bibinfo  {journal} {JHEP}\ }\textbf {\bibinfo {volume} {08}},\ \bibinfo {pages} {038} (\bibinfo {year} {2008})},\ \Eprint {http://arxiv.org/abs/0805.3165} {arXiv:0805.3165 [hep-ph]} \BibitemShut {NoStop}%
\bibitem [{\citenamefont {Meissner}\ \emph {et~al.}(2009)\citenamefont {Meissner}, \citenamefont {Metz},\ and\ \citenamefont {Schlegel}}]{Meissner:2009ww}%
  \BibitemOpen
  \bibfield  {author} {\bibinfo {author} {\bibfnamefont {S.}~\bibnamefont {Meissner}}, \bibinfo {author} {\bibfnamefont {A.}~\bibnamefont {Metz}}, \ and\ \bibinfo {author} {\bibfnamefont {M.}~\bibnamefont {Schlegel}},\ }\href {\doibase 10.1088/1126-6708/2009/08/056} {\bibfield  {journal} {\bibinfo  {journal} {JHEP}\ }\textbf {\bibinfo {volume} {08}},\ \bibinfo {pages} {056} (\bibinfo {year} {2009})},\ \Eprint {http://arxiv.org/abs/0906.5323} {arXiv:0906.5323 [hep-ph]} \BibitemShut {NoStop}%
\bibitem [{\citenamefont {Balitsky}(1996)}]{Balitsky:1995ub}%
  \BibitemOpen
  \bibfield  {author} {\bibinfo {author} {\bibfnamefont {I.}~\bibnamefont {Balitsky}},\ }\href {\doibase 10.1016/0550-3213(95)00638-9} {\bibfield  {journal} {\bibinfo  {journal} {Nucl. Phys. B}\ }\textbf {\bibinfo {volume} {463}},\ \bibinfo {pages} {99} (\bibinfo {year} {1996})},\ \Eprint {http://arxiv.org/abs/hep-ph/9509348} {arXiv:hep-ph/9509348} \BibitemShut {NoStop}%
\bibitem [{\citenamefont {Kovchegov}(1999)}]{Kovchegov:1999yj}%
  \BibitemOpen
  \bibfield  {author} {\bibinfo {author} {\bibfnamefont {Y.~V.}\ \bibnamefont {Kovchegov}},\ }\href {\doibase 10.1103/PhysRevD.60.034008} {\bibfield  {journal} {\bibinfo  {journal} {Phys. Rev. D}\ }\textbf {\bibinfo {volume} {60}},\ \bibinfo {pages} {034008} (\bibinfo {year} {1999})},\ \Eprint {http://arxiv.org/abs/hep-ph/9901281} {arXiv:hep-ph/9901281} \BibitemShut {NoStop}%
\bibitem [{\citenamefont {Jalilian-Marian}\ \emph {et~al.}(1998)\citenamefont {Jalilian-Marian}, \citenamefont {Kovner}, \citenamefont {Leonidov},\ and\ \citenamefont {Weigert}}]{Jalilian-Marian:1997jhx}%
  \BibitemOpen
  \bibfield  {author} {\bibinfo {author} {\bibfnamefont {J.}~\bibnamefont {Jalilian-Marian}}, \bibinfo {author} {\bibfnamefont {A.}~\bibnamefont {Kovner}}, \bibinfo {author} {\bibfnamefont {A.}~\bibnamefont {Leonidov}}, \ and\ \bibinfo {author} {\bibfnamefont {H.}~\bibnamefont {Weigert}},\ }\href {\doibase 10.1103/PhysRevD.59.014014} {\bibfield  {journal} {\bibinfo  {journal} {Phys. Rev. D}\ }\textbf {\bibinfo {volume} {59}},\ \bibinfo {pages} {014014} (\bibinfo {year} {1998})},\ \Eprint {http://arxiv.org/abs/hep-ph/9706377} {arXiv:hep-ph/9706377} \BibitemShut {NoStop}%
\bibitem [{\citenamefont {Iancu}\ \emph {et~al.}(2001)\citenamefont {Iancu}, \citenamefont {Leonidov},\ and\ \citenamefont {McLerran}}]{Iancu:2000hn}%
  \BibitemOpen
  \bibfield  {author} {\bibinfo {author} {\bibfnamefont {E.}~\bibnamefont {Iancu}}, \bibinfo {author} {\bibfnamefont {A.}~\bibnamefont {Leonidov}}, \ and\ \bibinfo {author} {\bibfnamefont {L.~D.}\ \bibnamefont {McLerran}},\ }\href {\doibase 10.1016/S0375-9474(01)00642-X} {\bibfield  {journal} {\bibinfo  {journal} {Nucl. Phys. A}\ }\textbf {\bibinfo {volume} {692}},\ \bibinfo {pages} {583} (\bibinfo {year} {2001})},\ \Eprint {http://arxiv.org/abs/hep-ph/0011241} {arXiv:hep-ph/0011241} \BibitemShut {NoStop}%
\bibitem [{\citenamefont {Balitsky}\ and\ \citenamefont {Chirilli}(2008)}]{Balitsky:2007feb}%
  \BibitemOpen
  \bibfield  {author} {\bibinfo {author} {\bibfnamefont {I.}~\bibnamefont {Balitsky}}\ and\ \bibinfo {author} {\bibfnamefont {G.~A.}\ \bibnamefont {Chirilli}},\ }\href {\doibase 10.1103/PhysRevD.77.014019} {\bibfield  {journal} {\bibinfo  {journal} {Phys. Rev. D}\ }\textbf {\bibinfo {volume} {77}},\ \bibinfo {pages} {014019} (\bibinfo {year} {2008})},\ \Eprint {http://arxiv.org/abs/0710.4330} {arXiv:0710.4330 [hep-ph]} \BibitemShut {NoStop}%
\bibitem [{\citenamefont {Boussarie}\ and\ \citenamefont {Mehtar-Tani}(2022)}]{Boussarie:2020fpb}%
  \BibitemOpen
  \bibfield  {author} {\bibinfo {author} {\bibfnamefont {R.}~\bibnamefont {Boussarie}}\ and\ \bibinfo {author} {\bibfnamefont {Y.}~\bibnamefont {Mehtar-Tani}},\ }\href {\doibase 10.1016/j.physletb.2022.137125} {\bibfield  {journal} {\bibinfo  {journal} {Phys. Lett. B}\ }\textbf {\bibinfo {volume} {831}},\ \bibinfo {pages} {137125} (\bibinfo {year} {2022})},\ \Eprint {http://arxiv.org/abs/2006.14569} {arXiv:2006.14569 [hep-ph]} \BibitemShut {NoStop}%
\bibitem [{\citenamefont {Boussarie}\ and\ \citenamefont {Mehtar-Tani}(2023)}]{Boussarie:2023xun}%
  \BibitemOpen
  \bibfield  {author} {\bibinfo {author} {\bibfnamefont {R.}~\bibnamefont {Boussarie}}\ and\ \bibinfo {author} {\bibfnamefont {Y.}~\bibnamefont {Mehtar-Tani}},\ }\href@noop {} {\  (\bibinfo {year} {2023})},\ \Eprint {http://arxiv.org/abs/2309.16576} {arXiv:2309.16576 [hep-ph]} \BibitemShut {NoStop}%
\bibitem [{\citenamefont {McLerran}\ and\ \citenamefont {Venugopalan}(1994{\natexlab{a}})}]{McLerran:1993ni}%
  \BibitemOpen
  \bibfield  {author} {\bibinfo {author} {\bibfnamefont {L.~D.}\ \bibnamefont {McLerran}}\ and\ \bibinfo {author} {\bibfnamefont {R.}~\bibnamefont {Venugopalan}},\ }\href {\doibase 10.1103/PhysRevD.49.2233} {\bibfield  {journal} {\bibinfo  {journal} {Phys. Rev. D}\ }\textbf {\bibinfo {volume} {49}},\ \bibinfo {pages} {2233} (\bibinfo {year} {1994}{\natexlab{a}})},\ \Eprint {http://arxiv.org/abs/hep-ph/9309289} {arXiv:hep-ph/9309289} \BibitemShut {NoStop}%
\bibitem [{\citenamefont {McLerran}\ and\ \citenamefont {Venugopalan}(1994{\natexlab{b}})}]{McLerran:1993ka}%
  \BibitemOpen
  \bibfield  {author} {\bibinfo {author} {\bibfnamefont {L.~D.}\ \bibnamefont {McLerran}}\ and\ \bibinfo {author} {\bibfnamefont {R.}~\bibnamefont {Venugopalan}},\ }\href {\doibase 10.1103/PhysRevD.49.3352} {\bibfield  {journal} {\bibinfo  {journal} {Phys. Rev. D}\ }\textbf {\bibinfo {volume} {49}},\ \bibinfo {pages} {3352} (\bibinfo {year} {1994}{\natexlab{b}})},\ \Eprint {http://arxiv.org/abs/hep-ph/9311205} {arXiv:hep-ph/9311205} \BibitemShut {NoStop}%
\bibitem [{\citenamefont {Jalilian-Marian}\ \emph {et~al.}(1997)\citenamefont {Jalilian-Marian}, \citenamefont {Kovner}, \citenamefont {McLerran},\ and\ \citenamefont {Weigert}}]{Jalilian-Marian:1996mkd}%
  \BibitemOpen
  \bibfield  {author} {\bibinfo {author} {\bibfnamefont {J.}~\bibnamefont {Jalilian-Marian}}, \bibinfo {author} {\bibfnamefont {A.}~\bibnamefont {Kovner}}, \bibinfo {author} {\bibfnamefont {L.~D.}\ \bibnamefont {McLerran}}, \ and\ \bibinfo {author} {\bibfnamefont {H.}~\bibnamefont {Weigert}},\ }\href {\doibase 10.1103/PhysRevD.55.5414} {\bibfield  {journal} {\bibinfo  {journal} {Phys. Rev. D}\ }\textbf {\bibinfo {volume} {55}},\ \bibinfo {pages} {5414} (\bibinfo {year} {1997})},\ \Eprint {http://arxiv.org/abs/hep-ph/9606337} {arXiv:hep-ph/9606337} \BibitemShut {NoStop}%
\bibitem [{\citenamefont {McLerran}\ and\ \citenamefont {Venugopalan}(1999)}]{McLerran:1998nk}%
  \BibitemOpen
  \bibfield  {author} {\bibinfo {author} {\bibfnamefont {L.~D.}\ \bibnamefont {McLerran}}\ and\ \bibinfo {author} {\bibfnamefont {R.}~\bibnamefont {Venugopalan}},\ }\href {\doibase 10.1103/PhysRevD.59.094002} {\bibfield  {journal} {\bibinfo  {journal} {Phys. Rev. D}\ }\textbf {\bibinfo {volume} {59}},\ \bibinfo {pages} {094002} (\bibinfo {year} {1999})},\ \Eprint {http://arxiv.org/abs/hep-ph/9809427} {arXiv:hep-ph/9809427} \BibitemShut {NoStop}%
\bibitem [{\citenamefont {Marquet}\ \emph {et~al.}(2009)\citenamefont {Marquet}, \citenamefont {Xiao},\ and\ \citenamefont {Yuan}}]{Marquet:2009ca}%
  \BibitemOpen
  \bibfield  {author} {\bibinfo {author} {\bibfnamefont {C.}~\bibnamefont {Marquet}}, \bibinfo {author} {\bibfnamefont {B.-W.}\ \bibnamefont {Xiao}}, \ and\ \bibinfo {author} {\bibfnamefont {F.}~\bibnamefont {Yuan}},\ }\href {\doibase 10.1016/j.physletb.2009.10.099} {\bibfield  {journal} {\bibinfo  {journal} {Phys. Lett. B}\ }\textbf {\bibinfo {volume} {682}},\ \bibinfo {pages} {207} (\bibinfo {year} {2009})},\ \Eprint {http://arxiv.org/abs/0906.1454} {arXiv:0906.1454 [hep-ph]} \BibitemShut {NoStop}%
\bibitem [{\citenamefont {Peskin}\ and\ \citenamefont {Schroeder}(1995)}]{Peskin:1995ev}%
  \BibitemOpen
  \bibfield  {author} {\bibinfo {author} {\bibfnamefont {M.~E.}\ \bibnamefont {Peskin}}\ and\ \bibinfo {author} {\bibfnamefont {D.~V.}\ \bibnamefont {Schroeder}},\ }\href@noop {} {\emph {\bibinfo {title} {{An Introduction to quantum field theory}}}}\ (\bibinfo  {publisher} {Addison-Wesley},\ \bibinfo {address} {Reading, USA},\ \bibinfo {year} {1995})\BibitemShut {NoStop}%
\bibitem [{\citenamefont {McLerran}(2002)}]{McLerran:2001sr}%
  \BibitemOpen
  \bibfield  {author} {\bibinfo {author} {\bibfnamefont {L.~D.}\ \bibnamefont {McLerran}},\ }\href {\doibase 10.1007/3-540-45792-5_8} {\bibfield  {journal} {\bibinfo  {journal} {Lect. Notes Phys.}\ }\textbf {\bibinfo {volume} {583}},\ \bibinfo {pages} {291} (\bibinfo {year} {2002})},\ \Eprint {http://arxiv.org/abs/hep-ph/0104285} {arXiv:hep-ph/0104285} \BibitemShut {NoStop}%
\bibitem [{\citenamefont {Metz}\ and\ \citenamefont {Zhou}(2011)}]{Metz:2011wb}%
  \BibitemOpen
  \bibfield  {author} {\bibinfo {author} {\bibfnamefont {A.}~\bibnamefont {Metz}}\ and\ \bibinfo {author} {\bibfnamefont {J.}~\bibnamefont {Zhou}},\ }\href {\doibase 10.1103/PhysRevD.84.051503} {\bibfield  {journal} {\bibinfo  {journal} {Phys. Rev. D}\ }\textbf {\bibinfo {volume} {84}},\ \bibinfo {pages} {051503} (\bibinfo {year} {2011})},\ \Eprint {http://arxiv.org/abs/1105.1991} {arXiv:1105.1991 [hep-ph]} \BibitemShut {NoStop}%
\bibitem [{\citenamefont {Bliokh}\ and\ \citenamefont {Nori}(2015)}]{Bliokh:2015doa}%
  \BibitemOpen
  \bibfield  {author} {\bibinfo {author} {\bibfnamefont {K.~Y.}\ \bibnamefont {Bliokh}}\ and\ \bibinfo {author} {\bibfnamefont {F.}~\bibnamefont {Nori}},\ }\href {\doibase 10.1016/j.physrep.2015.06.003} {\bibfield  {journal} {\bibinfo  {journal} {Phys. Rept.}\ }\textbf {\bibinfo {volume} {592}},\ \bibinfo {pages} {1} (\bibinfo {year} {2015})},\ \Eprint {http://arxiv.org/abs/1504.03113} {arXiv:1504.03113 [physics.optics]} \BibitemShut {NoStop}%
\bibitem [{\citenamefont {Kutak}(2011)}]{Kutak:2011rb}%
  \BibitemOpen
  \bibfield  {author} {\bibinfo {author} {\bibfnamefont {K.}~\bibnamefont {Kutak}},\ }\href {\doibase 10.1016/j.physletb.2011.09.113} {\bibfield  {journal} {\bibinfo  {journal} {Phys. Lett. B}\ }\textbf {\bibinfo {volume} {705}},\ \bibinfo {pages} {217} (\bibinfo {year} {2011})},\ \Eprint {http://arxiv.org/abs/1103.3654} {arXiv:1103.3654 [hep-ph]} \BibitemShut {NoStop}%
\bibitem [{\citenamefont {Kovner}\ and\ \citenamefont {Lublinsky}(2015)}]{Kovner:2015hga}%
  \BibitemOpen
  \bibfield  {author} {\bibinfo {author} {\bibfnamefont {A.}~\bibnamefont {Kovner}}\ and\ \bibinfo {author} {\bibfnamefont {M.}~\bibnamefont {Lublinsky}},\ }\href {\doibase 10.1103/PhysRevD.92.034016} {\bibfield  {journal} {\bibinfo  {journal} {Phys. Rev. D}\ }\textbf {\bibinfo {volume} {92}},\ \bibinfo {pages} {034016} (\bibinfo {year} {2015})},\ \Eprint {http://arxiv.org/abs/1506.05394} {arXiv:1506.05394 [hep-ph]} \BibitemShut {NoStop}%
\bibitem [{\citenamefont {Kharzeev}\ and\ \citenamefont {Levin}(2017)}]{Kharzeev:2017qzs}%
  \BibitemOpen
  \bibfield  {author} {\bibinfo {author} {\bibfnamefont {D.~E.}\ \bibnamefont {Kharzeev}}\ and\ \bibinfo {author} {\bibfnamefont {E.~M.}\ \bibnamefont {Levin}},\ }\href {\doibase 10.1103/PhysRevD.95.114008} {\bibfield  {journal} {\bibinfo  {journal} {Phys. Rev. D}\ }\textbf {\bibinfo {volume} {95}},\ \bibinfo {pages} {114008} (\bibinfo {year} {2017})},\ \Eprint {http://arxiv.org/abs/1702.03489} {arXiv:1702.03489 [hep-ph]} \BibitemShut {NoStop}%
\bibitem [{\citenamefont {Hagiwara}\ \emph {et~al.}(2018)\citenamefont {Hagiwara}, \citenamefont {Hatta}, \citenamefont {Xiao},\ and\ \citenamefont {Yuan}}]{Hagiwara:2017uaz}%
  \BibitemOpen
  \bibfield  {author} {\bibinfo {author} {\bibfnamefont {Y.}~\bibnamefont {Hagiwara}}, \bibinfo {author} {\bibfnamefont {Y.}~\bibnamefont {Hatta}}, \bibinfo {author} {\bibfnamefont {B.-W.}\ \bibnamefont {Xiao}}, \ and\ \bibinfo {author} {\bibfnamefont {F.}~\bibnamefont {Yuan}},\ }\href {\doibase 10.1103/PhysRevD.97.094029} {\bibfield  {journal} {\bibinfo  {journal} {Phys. Rev. D}\ }\textbf {\bibinfo {volume} {97}},\ \bibinfo {pages} {094029} (\bibinfo {year} {2018})},\ \Eprint {http://arxiv.org/abs/1801.00087} {arXiv:1801.00087 [hep-ph]} \BibitemShut {NoStop}%
\bibitem [{\citenamefont {Neill}\ and\ \citenamefont {Waalewijn}(2019)}]{Neill:2018uqw}%
  \BibitemOpen
  \bibfield  {author} {\bibinfo {author} {\bibfnamefont {D.}~\bibnamefont {Neill}}\ and\ \bibinfo {author} {\bibfnamefont {W.~J.}\ \bibnamefont {Waalewijn}},\ }\href {\doibase 10.1103/PhysRevLett.123.142001} {\bibfield  {journal} {\bibinfo  {journal} {Phys. Rev. Lett.}\ }\textbf {\bibinfo {volume} {123}},\ \bibinfo {pages} {142001} (\bibinfo {year} {2019})},\ \Eprint {http://arxiv.org/abs/1811.01021} {arXiv:1811.01021 [hep-ph]} \BibitemShut {NoStop}%
\bibitem [{\citenamefont {Dvali}\ and\ \citenamefont {Venugopalan}(2022)}]{Dvali:2021ooc}%
  \BibitemOpen
  \bibfield  {author} {\bibinfo {author} {\bibfnamefont {G.}~\bibnamefont {Dvali}}\ and\ \bibinfo {author} {\bibfnamefont {R.}~\bibnamefont {Venugopalan}},\ }\href {\doibase 10.1103/PhysRevD.105.056026} {\bibfield  {journal} {\bibinfo  {journal} {Phys. Rev. D}\ }\textbf {\bibinfo {volume} {105}},\ \bibinfo {pages} {056026} (\bibinfo {year} {2022})},\ \Eprint {http://arxiv.org/abs/2106.11989} {arXiv:2106.11989 [hep-th]} \BibitemShut {NoStop}%
\bibitem [{\citenamefont {Dumitru}\ \emph {et~al.}(2023)\citenamefont {Dumitru}, \citenamefont {Kovner},\ and\ \citenamefont {Skokov}}]{Dumitru:2023qee}%
  \BibitemOpen
  \bibfield  {author} {\bibinfo {author} {\bibfnamefont {A.}~\bibnamefont {Dumitru}}, \bibinfo {author} {\bibfnamefont {A.}~\bibnamefont {Kovner}}, \ and\ \bibinfo {author} {\bibfnamefont {V.~V.}\ \bibnamefont {Skokov}},\ }\href {\doibase 10.1103/PhysRevD.108.014014} {\bibfield  {journal} {\bibinfo  {journal} {Phys. Rev. D}\ }\textbf {\bibinfo {volume} {108}},\ \bibinfo {pages} {014014} (\bibinfo {year} {2023})},\ \Eprint {http://arxiv.org/abs/2304.08564} {arXiv:2304.08564 [hep-ph]} \BibitemShut {NoStop}%
\bibitem [{\citenamefont {Bhattacharya}\ \emph {et~al.}(2022{\natexlab{b}})\citenamefont {Bhattacharya}, \citenamefont {Boussarie},\ and\ \citenamefont {Hatta}}]{Bhattacharya:2022vvo}%
  \BibitemOpen
  \bibfield  {author} {\bibinfo {author} {\bibfnamefont {S.}~\bibnamefont {Bhattacharya}}, \bibinfo {author} {\bibfnamefont {R.}~\bibnamefont {Boussarie}}, \ and\ \bibinfo {author} {\bibfnamefont {Y.}~\bibnamefont {Hatta}},\ }\href {\doibase 10.1103/PhysRevLett.128.182002} {\bibfield  {journal} {\bibinfo  {journal} {Phys. Rev. Lett.}\ }\textbf {\bibinfo {volume} {128}},\ \bibinfo {pages} {182002} (\bibinfo {year} {2022}{\natexlab{b}})},\ \Eprint {http://arxiv.org/abs/2201.08709} {arXiv:2201.08709 [hep-ph]} \BibitemShut {NoStop}%
\bibitem [{\citenamefont {Bhattacharya}\ \emph {et~al.}(2024)\citenamefont {Bhattacharya}, \citenamefont {Boussarie},\ and\ \citenamefont {Hatta}}]{Bhattacharya:2024sck}%
  \BibitemOpen
  \bibfield  {author} {\bibinfo {author} {\bibfnamefont {S.}~\bibnamefont {Bhattacharya}}, \bibinfo {author} {\bibfnamefont {R.}~\bibnamefont {Boussarie}}, \ and\ \bibinfo {author} {\bibfnamefont {Y.}~\bibnamefont {Hatta}},\ }\href@noop {} {\  (\bibinfo {year} {2024})},\ \Eprint {http://arxiv.org/abs/2404.04209} {arXiv:2404.04209 [hep-ph]} \BibitemShut {NoStop}%
\end{thebibliography}%

\end{document}